\documentstyle[pra,aps,epsfig,multicol]{revtex}

\begin{document}

\title{Tailorable acceptor ${\rm C}_{60-n}{\rm B}_{n}$ and donor 
${\rm C}_{60-m}{\rm N}_{m}$ pairs for molecular electronics}

\author{ Rui-Hua Xie$^{1}$, Garnett W. Bryant$^{1}$, 
Jijun Zhao$^{2}$,  Vedene H. Smith, Jr.$^{3}$,  Aldo Di Carlo$^{4}$, 
and  Alessandro Pecchia$^{4}$ }
\address{$^{1}$Atomic Physics Division, National Institute 
of Standards and Technology, Gaithersburg, MD 20899-8423\\
$^{2}$ Department of Physics and Astronomy, University of North
Carolina at Chapel Hill, Chapel Hill, NC 27599\\
$^{3}$ Department of Chemistry, Queen's University, Kingston,
ON K7L 3N6, Canada\\ 
$^{4}$INFM-Dip. Ing. Elettronica, Universit\'{a} di Roma ``Tor Vergata", 
via del Politecnico 1, 00133 Roma, Italy}

\date{\today}

\maketitle

\begin{abstract}

Our first-principles calculations demonstrate that 
${\rm C}_{60-n}{\rm B}_{n}$ and ${\rm C}_{60-m}{\rm N}_{m}$ 
molecules can be engineered as the acceptors and  donors, respectively, 
which are needed for molecular electronics, by properly controlling the dopant 
number $n$ and $m$ in ${\rm C}_{60}$. As an example, we show that acceptor 
${\rm C}_{48}{\rm B}_{12}$ and donor ${\rm C}_{48}{\rm N}_{12}$ are promising 
components for molecular rectifiers, carbon nanotube-based $n$-$p$-$n$ 
($p$-$n$-$p$) transistors and  $p$-$n$ junctions.

{\bf PACS (numbers):} 33.15.-e, 85.65.+h, 31.15.Ar

\end{abstract}

\begin{multicols}{2} 

Modern microelectronics and computation are advancing at an extremely
fast rate because of remarkable circuit miniaturization\cite{berger97}.
However, this trend will soon reach the scale of atoms or molecules. 
To continue toward faster and smaller computers, new schemes are required. 
Molecular electronics\cite{aviram74} is one such approach. 

One major problem in molecular electronics is connecting the 
functional molecules and assigning the observed electrical properties in
an unambiguous way to the molecules in question\cite{taylor02,c60stm}. 
Fullerenes\cite{book1} are large enough to be identified by transmission 
electron microscopy or scanning probe methods\cite{c60stm}, 
 are stable and easy to build into molecular circuits 
\cite{book1,c60review}, and might be inserted into single-walled carbon 
nanotubes (SWNT)\cite{pichler01}. Hence, fullerenes should be ideal 
components for molecular electronics. 

As in semiconductor electronics\cite{berger97}, acceptor/donor pairs are 
critical  for use in molecular electronics, for example, molecular rectifiers 
\cite{aviram74}, nanoscale $p$-$n$-$p$ transistors and $p$-$n$ junctions\cite{pnjunc}. For 
traditional  silicon doping,  group V atoms (for example, phosphorous) 
act as donors and group  III ( for 
example, boron) are acceptors. Analogous acceptor/donor schemes are 
needed in molecular electronics\cite{aviram74}. 
Fullerenes are unique because they can be doped
in several different ways (for example,  endohedral\cite{book1}, 
substitutional\cite{book1,guo91,hummelen95,hultman01} and  exohedral doping\cite{book1}). 
This should provide a wide range of possible acceptor/donor schemes.

To design active molecular devices, components with controllable electronic properties
are needed\cite{aviram74,taylor02}. For
example,  to obtain molecular rectification, the lowest
unoccupied molecular orbital (LUMO) of the acceptor
should lie at or slightly above the Fermi level of the electrode and above
the highest occupied molecular orbital (HOMO) of the donor\cite{aviram74}.
Hence it is important to search for desired acceptor/donor pairs which satisfy 
the requirement. In this letter,  we suggest a controlled approach to
obtain such pairs from ${\rm C}_{60}$ molecules by using substitutional
doping. Because the average carbon-carbon bond length in 
${\rm C}_{60}$ is slightly larger than that in
graphite, which can only be substitutionally doped by boron,  and the force constants\cite{book1} are somewhat weakened
by the curvature of the ${\rm C}_{60}$ surface,  both boron  and nitrogen 
can substitute for one or more carbons in ${\rm C}_{60}$\cite{guo91,hummelen95,hultman01}. 
Our first-principles calculations  demonstrate that ${\rm C}_{60-n}{\rm B}_{n}$ and 
${\rm C}_{60-m}{\rm N}_{m}$ molecules can be engineered as the acceptors and donors, respectively, 
which are desired for molecular electronics by properly controlling the dopant number $n$ and $m$. 
 As an example, we present  the electronic  properties 
of the acceptor ${\rm C}_{48}{\rm B}_{12}$ and donor ${\rm C}_{48}{\rm N}_{12}$, 
and discuss their potential applications in molecular electronics.

First, we discuss the electronic properties of
${\rm C}_{60-n}{\rm B}_{n}$, ${\rm C}_{60-n}{\rm N}_{n}$
and ${\rm C}_{60}$ to show that ${\rm C}_{60-n}{\rm B}_{n}$ and 
${\rm C}_{60-n}{\rm N}_{n}$ can act as controlled dopants. The number $n$ ranges from 1 to 12. Since
there are many isomers\cite{isomer03} of ${\rm C}_{60-n}{\rm X}_{n}$ at fixed $n$,
we only take the dopant assignment of Hultman {\sl et al.} 
\cite{hultman01} as an example\cite{info03}: each pentagon receives a maximum of one 
dopant X, and two X are separated from each other by  two carbon 
atoms.  The optimized geometry and total energy are calculated by
using the Gaussian 98 program\cite{gaussian,nist} with the
B3LYP\cite{becke93} hybrid density functional theory (DFT)\cite{dft64} and
6-31G(d) basis set. The calculated results  are summarized
in Table I. It is found that the binding energy of ${\rm C}_{60-n}{\rm X}_{n}$ 
(X=B,N) decreases monotonically with increasing  integer $n$. 
For ${\rm C}_{60-n}{\rm B}_{n}$,  the binding energy is  
0.21 eV/atom to 0.38 eV/atom lower than that ($E_{b} = 6.98$ eV/atom) of ${\rm C}_{60}$. 
For ${\rm C}_{60-n}{\rm N}_{n}$, it is  0.10 eV/atom to  0.61 
eV/atom lower than that of ${\rm C}_{60}$. The stabilities of 
${\rm C}_{60-n}{\rm B}_{n}$  and ${\rm C}_{60-n}{\rm N}_{n}$ are comparable  
to but less than that of ${\rm C}_{60}$. In 1991, Smalley and coworkers\cite{guo91}
successfully synthesized ${\rm C}_{60-n}{\rm B}_{n}$ with $1 \le n \le 6$. 
In 1995, Hummelen {\sl et al.}\cite{hummelen95} produced 
${\rm C}_{59}{\rm N}$. Very recently, Hultman {\sl et al.}\cite{hultman01} have reported the
existence of ${\rm C}_{48}{\rm N}_{12}$. 
These experiments show that certain stable ${\rm C}_{60-n}{\rm X}_{n}$ (X=B, N) 
can be made. Our results show that other  B- or N-doped 
${\rm C}_{60-n}{\rm X}_{n}$ structures for $n\le 12$ 
have similar stability. Our 
calculated ionization potential  ($E_{I}= 7.32$ eV) and electron affinity 
($E_{A} = 2.40$ eV) for  ${\rm C}_{60}$ agree well with experiments (${\rm E}_{I} =
7.54\pm 0.01\ {\rm eV}$ \cite{hertel92} and ${\rm E}_{A} =
2.689 \pm 0.008\ {\rm eV}$ \cite{wang99}), indicating the accuracy of
our calculations for ${\rm C}_{60-n}{\rm X}_{n}$. Among
${\rm C}_{60}$ and  ${\rm C}_{60-n}{\rm X}_{n}$ (X=B,N) at fixed $n$,
${\rm C}_{60-n}{\rm B}_{n}$ has the highest electron affinity,
while ${\rm C}_{60-n}{\rm N}_{n}$ has the lowest
ionization potential, indicating that ${\rm C}_{60-n}{\rm B}_{n}$ and
${\rm C}_{60-n}{\rm N}_{n}$  can serve as  electron
acceptor and donor, respectively. 

Calculations of LUMO and HOMO energies are 
necessary to explore combinations of donors and acceptors suitable for 
molecular electronics. Table I  shows the change
of the LUMO/HOMO energies (and LUMO-HOMO gaps $E_{g}=E_{l}-E_{h}$)
as we vary the dopant number $n$ in the ${\rm C}_{60-n}{\rm X}_{n}$ molecule.
Hence,  acceptor/donor pairs which satisfy the required conditions in
molecular electronics can be obtained  by controlling the dopant number 
$n$ and $m$ in ${\rm C}_{60-n}{\rm B}_{n}$ and 
${\rm C}_{60-m}{\rm N}_{m}$.  For example, based on the acceptor LUMO and 
donor HOMO shown in Table I, we may choose acceptor/donor pairs, such as 
C$_{58}$B$_{2}$/C$_{58}$N$_{2}$, 
C$_{54}$B$_{6}$/C$_{54}$N$_{6}$, C$_{49}$B$_{11}$/C$_{51}$N$_{9}$, or 
C$_{48}$B$_{12}$/C$_{48}$N$_{12}$ to build molecular rectifiers exhibiting 
similar rectification behaviors. In the following, we take C$_{48}$B$_{12}$/C$_{48}$N$_{12}$ 
 as one example to show their applications in molecular electronics. 

\

\noindent
{\bf Table I:}  LUMO ($E_{l}$,  in eV), HOMO ($E_{h}$,
in eV),  binding energy ($E_{b}$,  in eV/atom), ionization potential
($E_{I}$,  in eV), and electron affinity  ($E_{A}$,  in eV) calculated for
 ${\rm C}_{60-n}{\rm X}_{n}$ (X=B,N)   using  B3LYP/6-31G(d).
 
\

\begin{center}
\begin{tabular}{cccccc|ccccc}\hline\hline
n    &\multicolumn{5}{c|}{${\rm C}_{60-n}{\rm B}_{n}$}
    &\multicolumn{5}{|c}{${\rm C}_{60-n}{\rm N}_{n}$} \\
   & $E_{l}$ & $E_{h}$ & $E_{b}$ & $E_{I}$ & $E_{A}$
    & $E_{l}$ & $E_{h}$ & $E_{b}$ & $E_{I}$ & $E_{A}$\\ \hline
1  &-4.26 & -5.57 & 6.77 & 6.77 & 3.07 &-3.63 & -4.72 &6.88  &5.93&2.15\\
2  &-4.64 & -5.25 & 6.75 & 6.35 & 3.53 &-3.71 &-4.64 &6.82 & 5.85 & 2.27  \\
3  &-4.21 & -5.49 & 6.74 & 6.37 & 3.32 & -3.72&-4.69 &6.78 & 5.92 &2.29\\
4  &-4.43 & -5.60 & 6.73 & 6.79 & 3.32 &-3.91 &-4.56 & 6.73  & 5.83 & 2.19 \\
5  &-4.59 & -5.54 & 6.71 & 6.85 & 3.56    &-3.74 & -4.56 & 6.69 &5.95&2.20\\
6  &-4.55  & -5.49  & 6.70 & 7.01 & 2.98 &-3.74 & -4.61 & 6.64 &5.85&2.32 \\
7  &-4.49 & -5.54 & 6.67 & 6.75 & 3.23 &-3.58 & -4.59 & 6.60&5.91&2.39 \\
8  &-4.62 & -5.34 & 6.66 & 6.44 & 3.50 &-3.60 & -4.41 & 6.55&5.65&2.36 \\
9  &-4.65 & -5.43 & 6.64 & 6.61 & 3.34 &-3.49 & -4.45 & 6.51&5.77&1.90 \\
10 &-4.57 & -5.44 & 6.63 & 6.57 & 3.45 &-3.31 & -4.28 & 6.46&5.52&1.89 \\
11 &-4.45 & -5.61 & 6.62 & 6.87 & 3.30 &-3.36 & -4.35 & 6.41&5.71&2.29 \\
12 &-4.24 & -5.58 & 6.60 & 6.73 & 3.08 &-2.61 & -4.38 & 6.37&5.66&1.49 \\ \hline
\end{tabular}
\end{center}

\
 
The molecular geometry (size) and symmetry play important roles in molecular
electronics \cite{aviram74,taylor02}. For example, an important factor in inducing rectification is some
geometric asymmetry in the molecular junction\cite{taylor02}. 
 ${\rm C}_{60}$ is a truncated icosahedron with a perfect
$I_{h}$ symmetry. Then, the changes in geometry (size) and symmetry of 
${\rm C}_{60-n}{\rm X}_{n}$, due to the dopant-induced effects, are needed 
to know.  Fig.1 shows the optimized structure of ${\rm C}_{48}{\rm B}_{12}$.
For comparison, similar calculations were done for  ${\rm C}_{60}$ and
${\rm C}_{48}{\rm N}_{12}$. The equilibrium ${\rm C}_{48}{\rm B}_{12}$, 
similar to ${\rm C}_{48}{\rm N}_{12}$\cite{hultman01},  has one dopant
 (boron) per pentagon  and  two dopants (boron) preferentially
sit in a hexagon. In  ${\rm C}_{48}{\rm X}_{12}$ (X=B, N),
the doping-induced distortion from the perfect ${\rm C}_{60}$ sphere is not
localized to the neighborhood of each dopant but extends
throughout the whole cage. The symmetry of ${\rm C}_{48}{\rm B}_{12}$ is
 the $C_{i}$ point group, different from the $S_{6}$ symmetry for
${\rm C}_{48}{\rm N}_{12}$\cite{hultman01}. This is  due to their
opposite electronic polarization, as discussed below.  The molecular structure
is examined by calculating the average distances (or radii ${\rm R}$) from
each atom to the center of
the molecule. ${\rm C}_{48}{\rm B}_{12}$ has an ellipsoidal structure, as
in ${\rm C}_{48}{\rm N}_{12}$, with 10 unique sites labelled 1 to 10 in 
Fig.1 and 10 different radii ranging from  0.347 nm (site 2, 2') to
0.387 nm (site 8, 8'). In contrast, the 10 unique radii for
${\rm C}_{48}{\rm N}_{12}$ range from 0.340 nm (site 7)
to 0.362 nm (site 5). For ${\rm C}_{60}$, each carbon atom has an equal
radius ${\rm R} = 0.355\ {\rm  nm}$, the same as that found in experiment
\cite{burgi92}. Furthermore, we find 15 unique bonds (specified in Fig.1)
in ${\rm C}_{48}{\rm B}_{12}$: 6 boron-carbon bonds with lengths between
0.154 nm and 0.159 nm, and 9 carbon-carbon bonds with 0.139 nm to 0.150 nm
lengths. In comparison, ${\rm C}_{48}{\rm N}_{12}$  has 
6 nitrogen-carbon bonds with lengths ranging from 0.141 nm to 0.143 nm, and
9 carbon-carbon bonds with lengths  from 0.139 nm to 0.145 nm. ${\rm C}_{60}$,
however, has one kind of single C-C  bond  (0.145 nm) and 
one kind of double  C=C bond (0.139 nm), which are in excellent agreement
with  experiments (0.14459 nm and 0.13997 nm \cite{burgi92}).

\begin{center}\epsfig{file=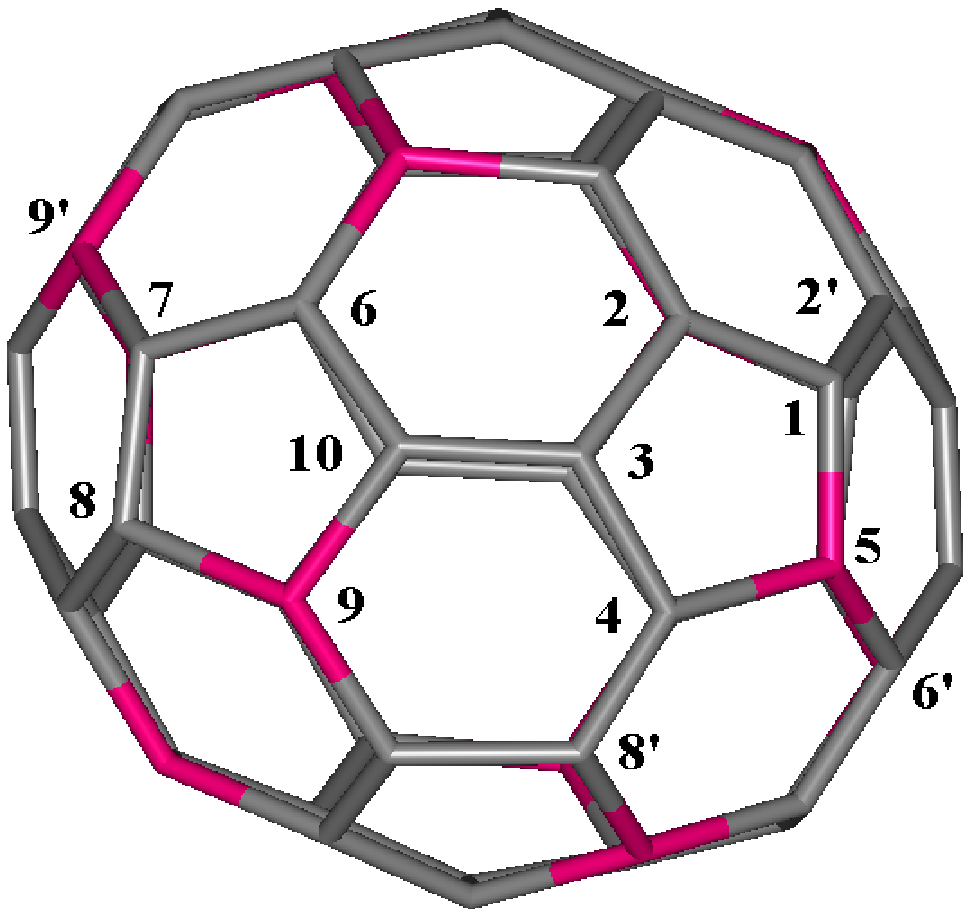,height=7cm,width=7cm}\end{center} 

{\bf FIG. 1.}:  ${\rm C}_{48}{\rm B}_{12}$ structure optimized with
B3LYP/6-31G(d). Red (grey) is for B (C) sites.
The 15 bonds between labeled vertices are all unique.

\

To determine the electronic polarization, Mulliken charge analysis
\cite{szabo82} was performed for the on-site charge ${\rm Q}_{m}$ of atoms in
${\rm C}_{48}{\rm B}_{12}$, ${\rm C}_{48}{\rm N}_{12}$  and ${\rm C}_{60}$.
${\rm C}_{48}{\rm B}_{12}$ has two types of boron dopants: one
boron with net Mulliken charge ${\rm Q}_{\rm m} = 0.164\ |e|$
(site 5) and one with ${\rm Q}_{\rm m} = 0.187\ |e|$ (site 9), while
carbon atoms in ${\rm C}_{48}{\rm B}_{12}$  have negative ${\rm Q}_{\rm m}$
in the range from -0.004 $|e|$ (site 3) to -0.083 $|e|$ (site 8).
In ${\rm C}_{48}{\rm N}_{12}$, there are two types of nitrogen dopants: one
nitrogen with ${\rm Q}_{\rm m} = -0.595\ |e|$  (site 5) and one with
${\rm Q}_{\rm m} = -0.600\ |e|$  (site 9), and two types of  carbon atoms:
one-fourth of the carbon atoms with negative ${\rm Q}_{\rm m}$ (-0.0125 $|e|$ at
site 2 and -0.0298 $|e|$ at site 3) and three-fourths of the carbon atoms with
${\rm Q}_{\rm m}$ in the range of 0.192 $|e|$ (site 6) to 0.227 $|e|$ (site 4).
In ${\rm C}_{60}$, ${\rm Q}_{\rm m}=0$  for each carbon atom.
Although the Mulliken analysis cannot estimate the atomic charges quantitatively,
their signs are correct\cite{szabo82}.  Hence, ${\rm C}_{60}$ is isotropic, 
while ${\rm C}_{48}{\rm B}_{12}$ and ${\rm C}_{48}{\rm N}_{12}$ have opposite
electronic polarizations, leading to their different symmetries.

\begin{center}\epsfig{file=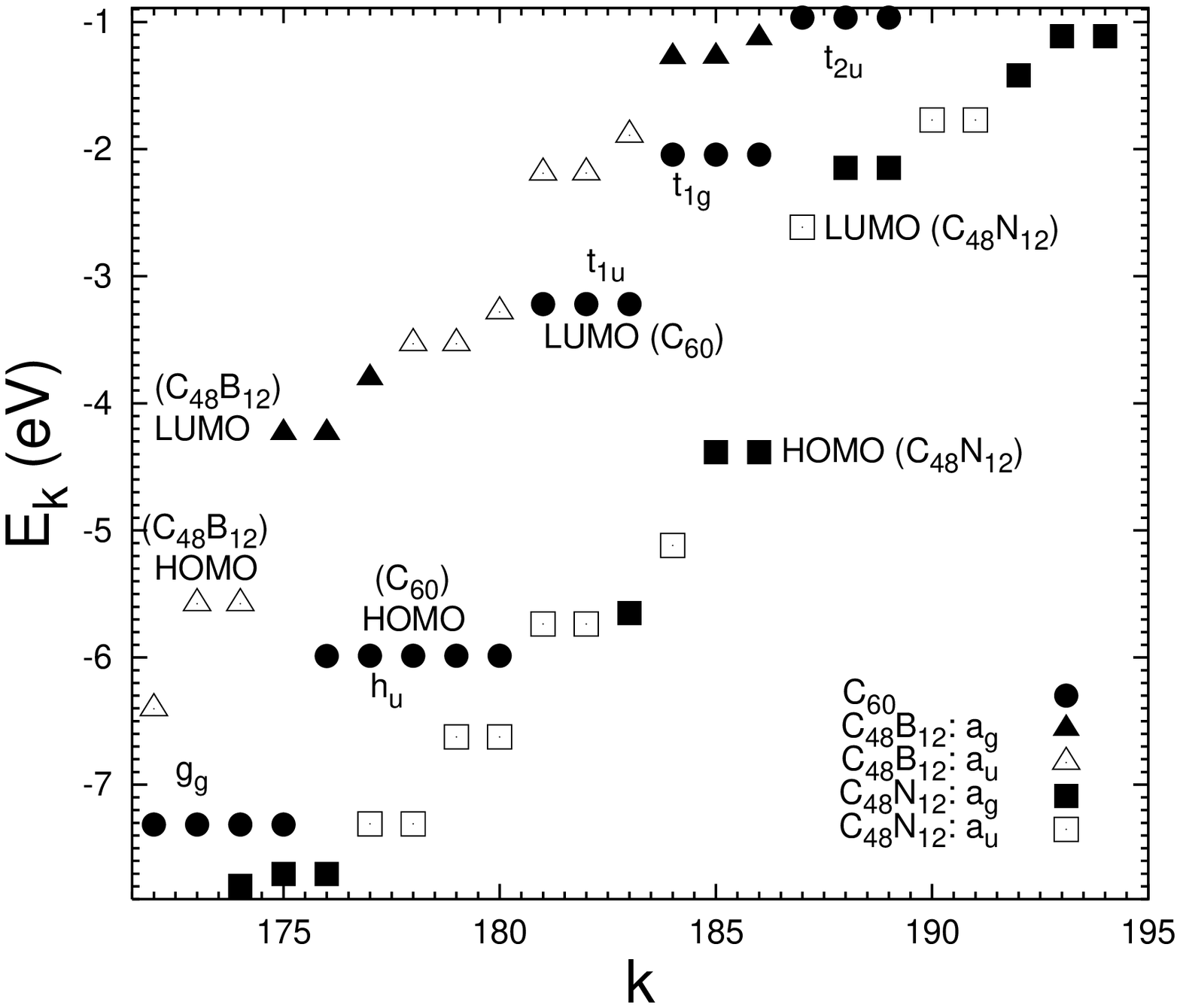,width=7cm,height=7cm}\end{center} 

{\bf  FIG. 2.}:  Orbital energies $E_{k}$ and symmetries of the $k$th eigenstate
for ${\rm C}_{48}{\rm X}_{12}$ (X=B,C,N) calculated with B3LYP/6-31G(d).

\

Applications of acceptors ${\rm C}_{60-n}{\rm B}_{n}$ and donors ${\rm C}_{60-n}{\rm N}_{n}$ 
require the proper line-up of energy levels. Upon boron or nitrogen doping, 
the degeneracy of the original ${\rm C}_{60}$ levels, shown in Fig.2, is 
removed by the structural distortion of the icosahedral symmetry. 
${\rm C}_{48}{\rm B}_{12}$ (${\rm C}_{48}{\rm N}_{12}$) is isoelectronic with 
${\rm C}_{60}^{+12}$ (${\rm C}_{60}^{-12}$). Replacing 12 carbon 
atoms by 12 boron atoms results in removing 10 electrons from the HOMO ($h_{u}$ 
symmetry, $E_{h} = -5.99$ eV) of ${\rm C}_{60}$  and 2 electrons from the 2nd HOMO ($g_{g}$ 
symmetry) of ${\rm C}_{60}$. In contrast, replacing 12 carbon atoms by 12 
nitrogen atoms leads to a complete filling of 6 electrons in the LUMO  ($t_{1u}$ symmetry, 
$E_{l} = -3.22$ eV) of ${\rm C}_{60}$ and 6 electrons  in the 2nd LUMO ($t_{1g}$ symmetry)  
of ${\rm C}_{60}$. Quantitatively,   Fig.2 shows the orbital energies of 
${\rm C}_{48}{\rm B}_{12}$ and ${\rm C}_{48}{\rm N}_{12}$. They are
different from each other  because of the difference in valency of B and N atoms. 
For ${\rm C}_{48}{\rm N}_{12}$, the HOMO is  a doubly-degenerate level 
of ${\rm a}_{\rm g}$ symmetry, while the LUMO is a non-degenerate level with 
${\rm a}_{\rm u}$ symmetry. For ${\rm C}_{48}{\rm B}_{12}$,  the HOMO and LUMO 
are doubly-degenerate levels of ${\rm a}_{\rm u}$ and ${\rm a}_{\rm g}$ 
symmetries,  respectively. These calculations show that ${\rm C}_{48}{\rm B}_{12}$'s LUMO 
 is just above ${\rm C}_{48}{\rm N}_{12}$'s HOMO  and  the approximate potential difference  
is about 1.63 eV. Thus, both ${\rm C}_{48}{\rm B}_{12}$ and ${\rm C}_{48}{\rm N}_{12}$  have 
precisely the properties required for a proper rectifier behavior as will be demonstrated 
below.

Various experiments\cite{rmm97} have shown that individual molecules can act as molecular rectifiers
\cite{aviram74}. The inset in Fig.3(a) shows a rectifier molecule  consisting of
a donor ${\rm C}_{48}{\rm N}_{12}$ and a acceptor 
${\rm C}_{48}{\rm B}_{12}$ connected by a tunneling bridge (a 
$\sigma$-electron system,  $-({\rm C}{\rm H}_{2})_{6}-$). The 
bridge forms a potential barrier that isolates the donor  from the acceptor 
on the time-scale of electron motion  to or from the electrodes\cite{aviram74} 
(see review \cite{c60review} about synthesis techniques which have successfully 
bridged ${\rm C}_{60}$ dimers through electroactive spacers).
Following the Aviram-Ratner scheme\cite{aviram74}, we expect that electron current would pass from 
the cathode to the anode via acceptor,  bridge and donor.  
Since  ${\rm C}_{48}{\rm B}_{12}$'s LUMO is -4.2 eV, we may choose 
metallic carbon nanotubes (work function $\phi\approx 5$ eV\cite{zhao02}),
gold, silver or copper ( $\phi\approx 4.4, 4.7, 4.8$  eV, respectively \cite{cnt02}) as the electrodes.
As discussed before, ${\rm C}_{48}{\rm B}_{12}$ and 
${\rm C}_{48}{\rm N}_{12}$  have  the properties required for 
proper rectifier behavior. An initial B3LYP/3-21G calculation for the full, 
connected fullerene/spacer/fullerene system shows that the covalent bond between the 
spacer and doped fullerenes shifts the electronic structure of the doped fullerenes. 
However, the HOMO and LUMO of the accepter-spacer-donor system are still localized 
on the acceptor and donor sides, respectively, ensuring the desired rectifier behavior. 
The same properties hold for the acceptor/bridge/donor system\cite{complex}. Thus, a 
smaller threshold voltage for conduction is expected in one direction than 
in  the other direction.   As an example, we have investigated the 
rectifying characteristic of the ${\rm C}_{48}{\rm B}_{12}/{\rm C}_{6}{\rm H}_{14}/{\rm C}_{48}{\rm N}_{12}$ 
system by using non-equilibrium Green function theory in conjunction with
a density functional based tight-binding model\cite{dicarlo,elstner}.
The rectifier molecule is connected to gold contacts that we treat in the
$s$-wave approximation with a constant density of states near to the
Fermi level\cite{dattaprl}. The calculated current/applied potential
characteristic is shown in Fig.3(a). A typical rectification
characteristic is obtained with a turn-on bias close to the
energy difference between  ${\rm C}_{48}{\rm B}_{12}$'s LUMO
and  ${\rm C}_{48}{\rm N}_{12}$'s HOMO. Recently, Joachim {\sl et al.}\cite{c60stm} 
have successfully observed the electrical current  flowing through an
individual ${\rm C}_{60}$ molecule with a scanning
tunneling microscope (STM). In light of this progress and the calculated size of 
${\rm C}_{48}{\rm X}_{12}$ (X=N,B), our rectifier molecule
also should be identifiable by STM. 

Heterojunctions for molecular electronics can also be formed using ${\rm C}_{60-n}{\rm X}_{n}$ 
donors and acceptors. Hybrid nanostructures formed by filling  a SWNT with
${\rm C}_{60}$\cite{pichler01}  have been observed, and can be 
superconducting \cite{pichler01} or metallic\cite{okada}. Our DFT calculations, done within the local 
density approximation with double numerical basis including 
d-polarization function\cite{nist,dmol}, show that
 the incorporation of the C$_{48}$B$_{12}$ or C$_{48}$N$_{12}$ 
into a (10,10) \cite{book1}  or (17,0)\cite{book1} 
 SWNT is energetically favorable. About 2.4 eV binding energy per molecule
 is gained after the C$_{48}$B$_{12}$ or C$_{48}$N$_{12}$ molecule 
is inserted periodically inside the SWNTs. In our calculations of doped 
SWNTs with a one-dimensional periodic boundary condition along the tube axis, 
three unit cells of a (10,10) tube or two unit cells of a (17,0) tube 
are included in the supercell with one ${\rm C}_{48}{\rm B}_{12}$ or 
${\rm C}_{48}{\rm N}_{12}$. Charge analysis found that 
placing an acceptor ${\rm C}_{48}{\rm B}_{12}$ into a (17,0) tube puts 
a +0.67 $|e|$ charge on the SWNT, while incorporating a donor 
${\rm C}_{48}{\rm N}_{12}$ into a (17,0) SWNT puts a -0.39 $|e|$
charge on the SWNT.  Similar results are obtained for a (10,10) SWNT. Hence,
putting ${\rm C}_{48}{\rm B}_{12}$ into a semiconducting tube results in  
a $p$-type region on the SWNT, while filling of donors 
${\rm C}_{48}{\rm N}_{12}$ into a  semiconducting  tube leads to a 
$n$-type region on the SWNT. Thus, it is possible to use them to get 
$n$-$p$-$n$ and $p$-$n$-$p$ transistors\cite{pnjunc}. 
As shown in Fig.3(b), doping a (17,0) semiconducting SWNT with ${\rm C}_{48}{\rm B}_{12}$ 
and ${\rm C}_{48}{\rm N}_{12}$ should also make a SWNT-based $p$-$n$ junction.

\

\centerline{\epsfxsize=3.5in \epsfbox{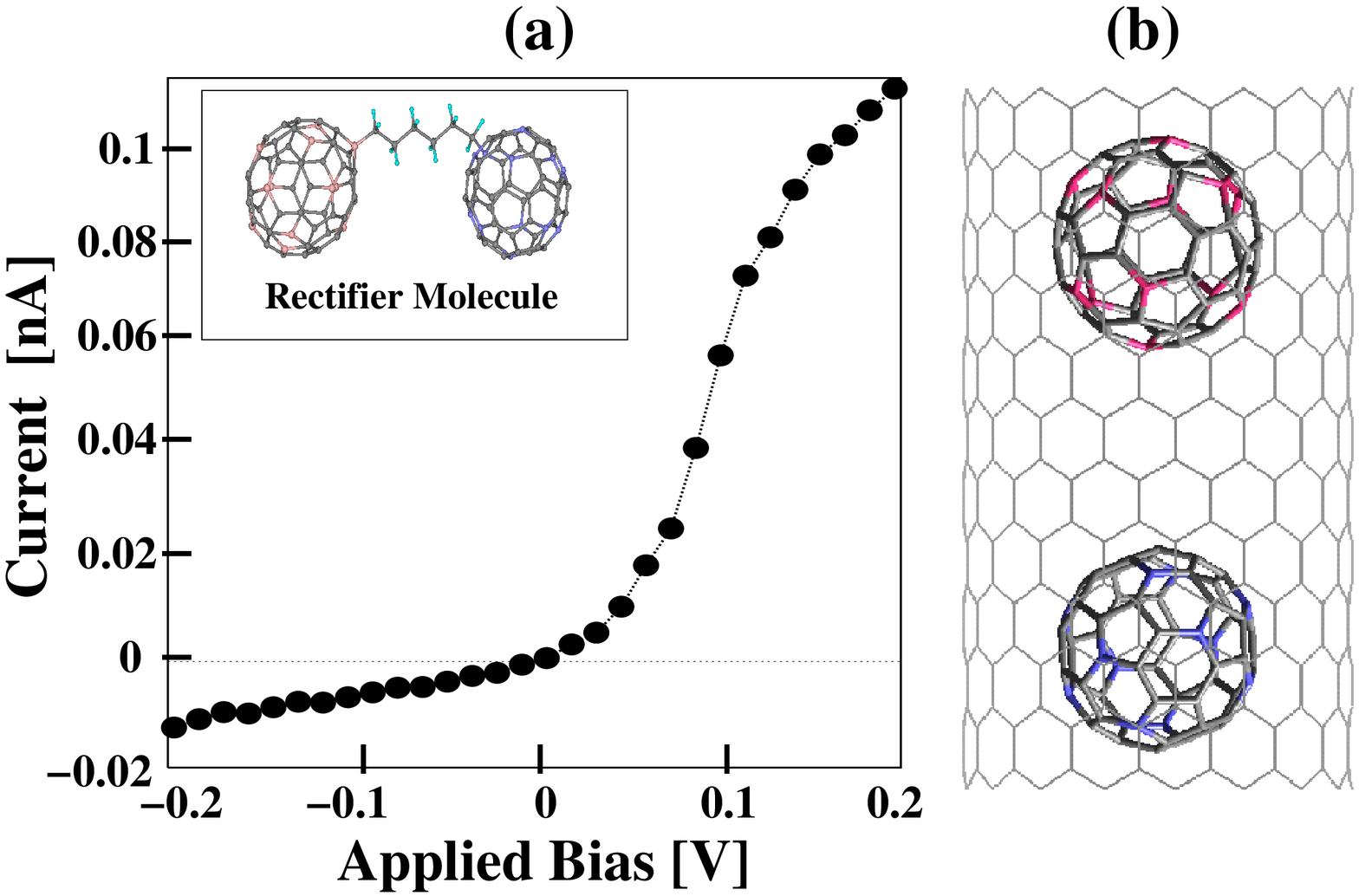}}
 
{\bf  FIG.3}:  (a) Calculated current through a rectifier molecule (inset)
which consists of ${\rm C}_{48}{\rm B}_{12}$ (left, red for
B atoms) and ${\rm C}_{48}{\rm N}_{12}$ (right, blue  for
N atoms)  connected by  ${\rm C}_{6}{\rm H}_{14}$ (middle,
green for H atoms) and is coupled to two Au electrodes via S
atoms on the two sides. The two C atoms in the ends of the
${\rm C}_{6}{\rm H}_{14}$ bridge are connected to the N (B)
atoms in ${\rm C}_{48}{\rm N}_{12}$ (${\rm C}_{48}{\rm B}_{12}$)
 respectively. A B3LYP/3-21G calculation shows that the connected rectifier
molecular superstructure is stable relative to the separated, individual components by
about 19.545 kcal/mol energy. (b) a prototype for ${\rm C}_{48}{\rm X}_{12}$@(17,0)
SWNT-based (X=B,N) $p$-$n$ junction.

\

In conclusion,  we show that acceptor 
${\rm C}_{60-n}{\rm B}_{n}$ and donor ${\rm C}_{60-m}{\rm N}_{m}$ pairs needed 
for molecular electronics can be obtained 
by properly controlling the number $n$ and $m$ of the substitutional dopants in 
${\rm C}_{60}$. 
We demonstrated the rectifying characteristic of 
a molecular rectifier built from our engineered acceptor/donor pairs. 
Heterojunctions for molecular electronics can be made by inserting these dopants 
into semiconducting carbon nanotubes. Efficiently synthesizing those acceptor/donor 
pairs would be of great experimental interest within reach of today's technology. 
Very recently, small nitrogen-substitutionally-doped  fullerenes have been reported, 
\cite{schultz03} showing experimental progress in this direction.

One of us (R.H.X.) thanks the HPCVL at Queen's University for the use of its parallel 
supercomputing facilities. 
 VHS gratefully acknowledges support from NSERC. ADC acknowledges 
support from the European DIODE Network and the Italian MURST-CNR under the project 
``Nanoelectronics".

\end{multicols}
 

\begin{references}
\bibitem{nist}  Use of this software does not constitute an endorsement
or certification by NIST.
\end{references}
\end{document}